\documentclass{article}
\usepackage{graphicx}
\begin{document}
\title{Metamorphic Relation Generation: State of the Art and Visions for Future Research}

\author{Rui Li\\	Deakin University\\ kevin.li@deakin.edu.au\\
		\and
		Huai Liu\\ Swinburne University of Technology \\ hliu@swin.edu.au\\
		\and
		Pak-Lok Poon\\ Central Queensland University \\ p.poon@cqu.edu.au
		\and
		Dave Towey\\ University of Nottingham Ningbo China \\ Dave.Towey@nottingham.edu.cn\\
		\and
		Chang-Ai Sun\\ University of Science and Technology Beijing\\ casun@ustb.edu.cn\\
		\and
		Zheng Zheng\\ Beihang University\\ zhengz@buaa.edu.cn\\
		\and
		Zhi Quan Zhou\\ University of Wollongong\\ zhiquan.zhou@gmail.com\\
		\and
		Tsong Yueh Chen\\ Swinburne University of Technology\\ tychen@swin.edu.au\\
	}
\date{}
\maketitle

\begin{abstract}
Metamorphic testing has become one mainstream technique to address the notorious oracle problem in software testing, thanks to its great successes in revealing real-life bugs in a wide variety of software systems. Metamorphic relations, the core component of metamorphic testing, have continuously attracted research interests from both academia and industry. In the last decade, a rapidly increasing number of studies have been conducted to systematically generate metamorphic relations from various sources and for different application domains. In this article, based on the systematic review on the state of the art for metamorphic relations' generation, we summarize and highlight visions for further advancing the theory and techniques for identifying and constructing metamorphic relations, and discuss potential research trends in related areas.
\end{abstract}

\maketitle

\section{Introduction}
\label{sec:introduction}

Software testing is a main approach for software quality assurance and control. A typical testing process involves analyzing the test object; designing, generating and executing test cases; and verifying test results. Various techniques have been developed to generate test cases, such as random testing~\cite{Ciupa2011OnTN}, partition testing~\cite{Ostrand88}, and fault-based testing~\cite{Morell90}, aiming at improving the effectiveness in detecting software faults. Theoretically speaking, a systematic mechanism, called the \textsl{test oracle}~\cite{Pezze14}, is needed to verify the execution result of each test case: If the test result is not consistent with the expected output, a failure is said to be revealed, which indicates the detection of software fault(s). However, in many practical situations, the oracle either does not exist or is extremely difficult to apply. Such an \textsl{oracle problem}~\cite{Barr15} significantly hinders the applicability and effectiveness of many testing techniques: If the correctness of the test result cannot be verified, we cannot judge whether the test case passes or fails the software under test. A few techniques have been proposed to address the oracle problem by replacing the oracle with some other test result verification mechanisms, such as \textsl{metamorphic testing} (MT)~\cite{Chen98, segura2016survey, chen2018metamorphic}  as well as those based on program assertions~\cite{terragni2020evolutionary} and multiple versions~\cite{securityNversion}.

Among the techniques for alleviating the oracle problem, MT has received much attention from the testing community in the last decade. In essence, applying MT is based on a set of necessary properties of the target software, which are represented in the form of \textsl{metamorphic relations} (MRs) among multiple program inputs and their corresponding expected outputs. Original test cases, called the \textsl{source test cases}, are generated by some testing techniques or selected from an existing test pool. MRs are then used to transform these source test cases into some new test cases, called the \textsl{follow-up test cases}. After executing both source and follow-up test cases, the multiple execution results are verified together against their corresponding MRs, instead of checking each individual test case's output, which may be infeasible due to the oracle problem. Since its invention in late 1990's~\cite{Chan1998ApplicationOM, Chen98}, MT has been successfully used to detect real-life bugs in a wide variety of software systems from different application domains, such as compilers~\cite{Le14, dreamMR21}, autonomous driving~\cite{Tian18, RMT23}, search engines~\cite{Zhou16}, machine learning classifiers~\cite{xie2011testing, Dwarakanath18}, and scientific simulations~\cite{LuuMET22}. MT was recently acknowledged as ``the most popular'' testing technique for AI-based systems~\cite{SE4AIsurvey}, and is the only new testing method added into an ISO/IEC/IEEE standard in the past two decades~\cite{ISO291194}. 

A key factor for the effective application of MT is a set of MRs, which supports not only test case generation but also test result verification. In the first decade after the introduction of MT, most MT-related studies only involved MRs that were identified in an \textit{ad hoc} manner. Although these MRs already demonstrated high fault detection capability, their ad hoc identification still has intrinsic shortcomings: Since the identification exercise largely depends on an individual tester's knowledge, skills and experience, it may be challenging to generate a sufficient number of high-quality MRs~\cite{GroceTSE14}. This problem has inspired a growing number of investigations into the systematic generation of MRs. For example, some studies focus on deriving new MRs based on existing ones~\cite{dong2008case, liu2012new, qiu2020theoretical}; others attempt to construct MRs from scratch~\cite{chen2016metric, sun2019metric}; still others make use of machine learning techniques to obtain valid MRs~\cite{kanewala2013using, kanewala2014techniques, kanewala2016predicting}. MR generation has recently emerged as one of the most important and popular topics in MT research. This is evident by the rapidly increasing publications on MR generation in the last decade. This trend is particularly profound in the last five years: 
Until December 31st 2023, there have been 75 papers on the systematic generation of MRs, among which 57 were published between 2019 and 2023.
Although there exist some reviews~\cite{segura2016survey, chen2018metamorphic} on the general context of MT, the literature lacks a study exclusively on MR generation, especially on its most up-to-date research outcomes and the new visions for the future work in the relevant areas: This is the main focus of this article.

\section{Background}
\label{sec:background}

MRs are the core component of MT: They are the basis for both test case generation and test result verification in MT. Let us look at an example to illustrate how MRs work when implementing MT. Suppose that a program $P$ calculates the shortest path between two nodes $a$ and $b$ for an undirected graph $G$. We can define an MR for $P$, termed as $\mathit{MR_{swap}}$: $|P(G, a, b)| = |P(G, b, a)|$, where $|\bullet|$ refers to the length of a path. For $\mathit{MR_{swap}}$, we have the source test case $\{G, a, b\}$ and the follow-up test case $\{G, b, a\}$. After executing them on the program $P$, we check whether the relation holds; if not, we say that $\mathit{MR_{swap}}$ is violated, indicating the detection of a fault.

It should be pointed out that MRs can have various forms. MRs are not limited to the style of one source and one follow-up test cases (like $\mathit{MR_{swap}}$). For example, for a program $\mathit{MM}$ calculating the product of two matrices, an MR, namely $\mathit{MR_{leftDist}}$, can be defined based on the property of ``left distributivity'' for matrix multiplication~\cite{Liu14TSE}: $A\times B + A\times C = A\times (B+C)$, where $A$ is an $m\times p$ matrix, and $B$ and $C$ are $p\times n$ matrices ($m, p, n = 1, 2, \ldots$). Based on the above definition, we can say that $\mathit{MR_{leftDist}}$ requires two source test cases $\{A, B\}$ and $\{A, C\}$, but only one follow-up test case $\{A, B+C\}$. 

The majority of existing MRs are simply composed of the \textit{input sub-relation} (the relation involving inputs only) and \textit{output sub-relation} (the relation involving outputs only). However, there also exists another type of MRs, where the construction of follow-up test cases requires the execution results of source test cases, as shown in the following example. Suppose that a search engine $S$ (like Google and Bing) can return a list of results based on a set of keywords. One possible MR for $S$, termed as $\mathit{MR_{site}}$~\cite{Zhou16}, can be defined as follows: Given the keywords $\{w_1, w_2, \ldots, w_n\}$, suppose that $S(w_1\,\mathrm{AND}\,w_2\,\mathrm{AND}\,\cdots\,\mathrm{AND}\,w_n)$ contains a non-empty list $L_d$ of results from a specific domain $d$ (such as $d = \texttt{.au}$ and $d = \texttt{.uk}$, referring to sites in Australia and UK, respectively). $L_d$ should also appear in the result $S(w_1\,\mathrm{AND}\,w_2\,\mathrm{AND}\,\cdots\,\mathrm{AND}\,w_n$; \texttt{site}:$d)$, where \mbox{``; \texttt{site}:$d$''} means that the search results will be limited to the domain $d$. For constructing the follow-up test case for $\mathit{MR_{site}}$, we need to first obtain the execution result of the source test case to get the value for $d$; in other words, the source test case must be executed before the follow-up test case can be constructed.

As clarified in previous studies~\cite{chen2018metamorphic}, MRs are not limited to equality relations. For example, in $\mathit{MR_{site}}$, the relation among the search results is actually ``subset'', that is, $S(w_1\,\mathrm{AND}\,w_2\,\mathrm{AND}\,\cdots\,\mathrm{AND}\,w_n$; \texttt{site}:$d) \subseteq S(w_1\,\mathrm{AND}\,w_2\,\mathrm{AND}\,\cdots\,\mathrm{AND}\,w_n)$.

Previous surveys on MT~\cite{Segura17TSE, chen2018metamorphic} have consistently recognized that systematic MR generation is a pressing  topic in MT research. Numerous studies have been conducted to propose a series of approaches or techniques to generate MRs for different application domains. Despite the importance of MR generation and the growing number of related studies, there does not exist a systematic study to provide a holistic view on MR generation and thus to facilitate further research on this area. These knowledge gaps will be addressed in this article, particularly in the following Sections~\ref{sec:reviews} and~\ref{sec:ResearchTrends}.


\section{State of the Art of Metamorphic Relations' Generation}
\label{sec:reviews}

To facilitate the systematic investigation of MR generation, we collected research articles published between January 1st 1998 (the year MT was introduced) and December 31st 2023, from five mainstream publishers including IEEE, ACM, Elsevier, Springer and Wiley. According to our research purpose, search keywords ``metamorphic testing'' and ``metamorphic relation'' were defined.
After obtaining an initial search pool, we scanned the abstract and introduction of each paper and even checked the main text of some papers in doubt, consequently removing papers irrelevant to MR generation. There exist some studies where MRs were manually generated in an ``ad hoc'' manner. They were also eliminated from our study. We finally obtained 75 papers that introduce certain techniques for systematically generating MRs, which compose the main investigation objects of this study.

\begin{figure}[t]
	\centering
	\includegraphics[width=\textwidth]{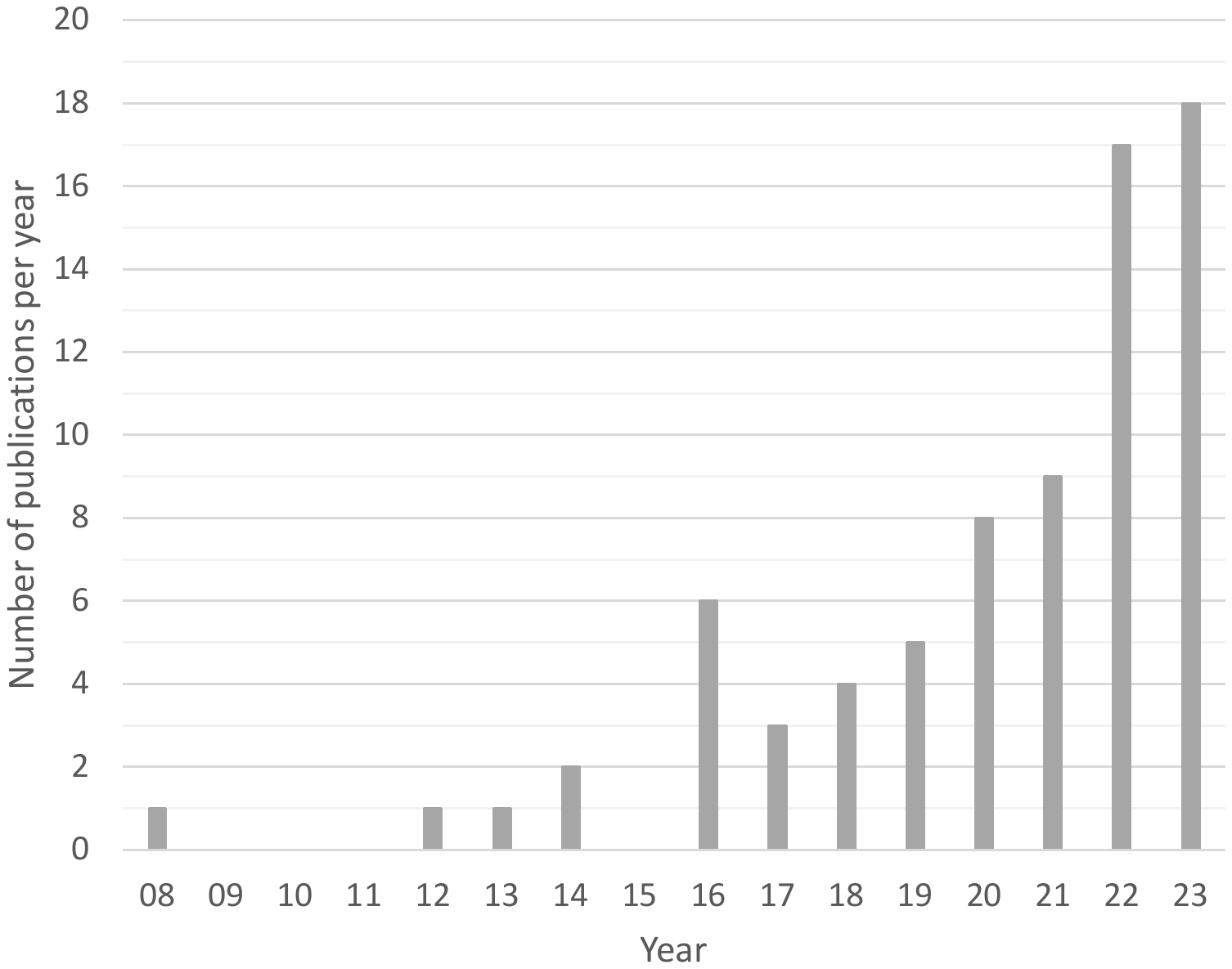}
	\caption{Number of publications on systematic MR generation per year}
	\label{fig:pb_year}
\end{figure}
\begin{figure}[t]
	\centering
	\includegraphics[width=\textwidth]{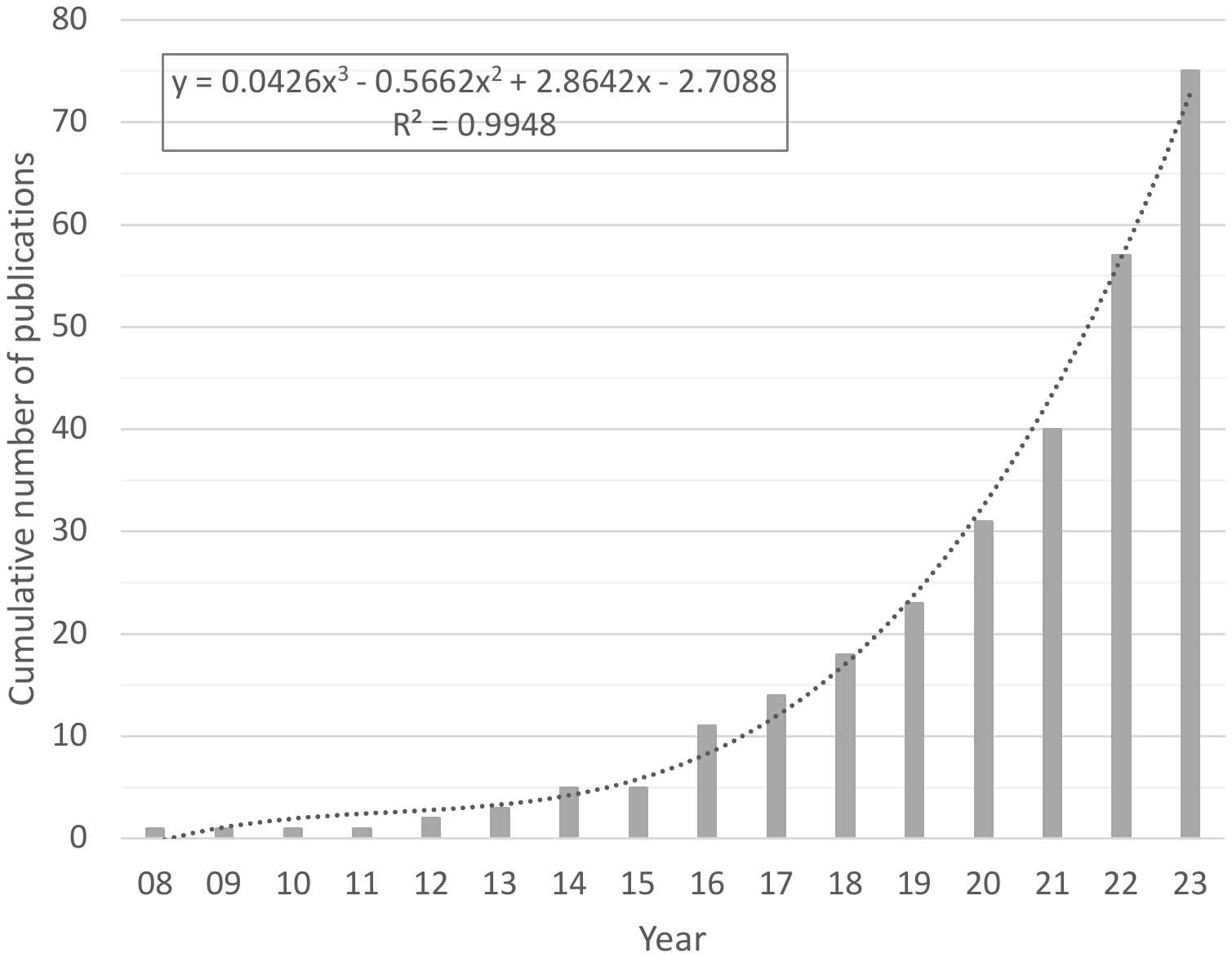}
	\caption{Cumulative number of publications on systematic MR generation}
	\label{fig:pb_cumu}
\end{figure}

It is interesting to observe the publication trend of systematic MR generation, as shown in Figs.~\ref{fig:pb_year} and~~\ref{fig:pb_cumu}, which illustrates the number of publications per year. Note that the first paper about systematic generation of MRs appeared in 2008~\cite{dong2008case}, ten years after the invention of MT. Thus, Figs.~\ref{fig:pb_year} and~~\ref{fig:pb_cumu} start from 2008. Fig.~\ref{fig:pb_year} illustrates the number of publications on systematically generating MRs per year. It can be observed that almost all of the papers were published in the last decade (only three papers were published in 2008, 2012 and 2013).
Fig.~\ref{fig:pb_cumu} gives the cumulative number of publications. 
To approximate the publication trend, we drew a trendline in the form of cubic polynomial function $0.0426x^3-0.5662x^2+2.8642x-2.7088$, with a very high determination coefficient $R^2=0.9948$. 
Following this trend, we anticipate that there would be a total of 280$+$ publications on systematic MR generation by 2030.

We coarsely classified the MR generation techniques into two main categories, either from existing MRs or from scratch, which will be reviewed in Sections~\ref{sec:composition} and~\ref{sec:ML}-\ref{sec:others}, respectively.

\subsection{Composition}
\label{sec:composition}

One of the earliest research directions for MR generation approaches is to construct new MRs by composing existing ones. By nature, MR composition can be simply implemented as follows. Given two MRs (also called \textit{component MRs}), $MR_1$ and $MR_2$, if any follow-up test case generated based on $MR_1$ can be used as the source test case for $MR_2$, then a new \textsl{composite MR}, i.e., $MR_{12}$, can be constructed. Note that (1) $MR_{21}$ and $MR_{12}$ normally refer to different relations; (2) there exist some situations where it is possible to obtain $MR_{12}$ but not $MR_{21}$ (or obtain $MR_{21}$ but not $MR_{12}$); and (3) some MRs cannot be composed with each other.

The first investigation into MR composition was conducted by Dong et al.~\cite{dong2008case}. Their empirical studies showed that the fault-detection effectiveness of composite MRs were dependent on corresponding individual component MRs as well as the composition sequence. It was also observed that a composite MR seemed better than its corresponding component MRs, in that the former could detect the faults which were collectively revealed by the latter.

Independent from Dong et al.'s work, Liu et al.~\cite{liu2012new} investigated the cost-effectiveness of composite MRs. They first gave a formal definition of the composability of MRs, and then conducted a series of experiments to compare composite and component MRs. The experimental results showed that using composite MRs was normally more cost-effective than using component MRs separately. The investigation also pointed out some plausible interpretations for when composite MRs would have better or poorer fault-detection capabilities than their component counterparts. 

More recently, Qiu et al.~\cite{qiu2020theoretical} conducted a comprehensive study, both theoretically and empirically, on the fault-detection effectiveness of composite MRs. The most significant contribution of their study is the provision of some definite answers about the sufficient conditions under which composite MRs must be more effective than component MRs. They also explained why composite MRs seemed better under most scenarios, and gave an easy-to-follow guideline that can help testers determine when they should use composite MRs instead of the corresponding component MRs.

\subsection{Artificial Intelligence Based Techniques}
\label{sec:ML}

Kanewala is the pioneer in using machine learning techniques for MR generation. The first work in this context was presented in 2013~\cite{kanewala2013using}, where machine learning was used to ``predict'' likely MRs. This approach consists of the following three main steps: (1) construct the control flow graph (CFG) for a target program/function; (2) extract features (including node and path features) from the CFG, based on which machine learning can create a predictive model; and (3) use the predictive model to predict which relations (from a set of pre-defined relations) are the likely MRs valid for the target program/function. Kanewala and Bieman~\cite{kanewala2013using} applied two machine learning techniques, namely the decision tree (DT) and support vector machine (SVM), to predict MRs for 48 scientific numeric functions (such as average, variation, and Manhattan distance). It was observed that the predicted MRs were effective in killing the majority of mutants of these functions, and SVM seemed superior to DT in mutant-killing capability. In addition to CFG, Kanewala~\cite{kanewala2014techniques} further suggested the program dependency graph (PDG) as the basis for extracting features. Also proposed was the utilization of two graph kernels, namely the random walk kernels and graphlet kernels, for comparing ``different graph substructures'' in order to obtain some useful information for identifying MRs.

On top of the above studies~\cite{kanewala2013using, kanewala2014techniques}, Kanewala et al.~\cite{kanewala2016predicting} conducted a more comprehensive study, where they used the graph kernel techniques (both random walk and graphlet kernels) to measure the similarity between program structures and thus to extend the features extracted from CFG/PDG. They found that the graph kernels did help improve the accuracy of MR prediction. They also observed that the control flow information was normally more effective than the data dependency information in predicting MRs. The applicability and effectiveness of this method were further justified by other experimental studies, such as those conducted by Rahman and Kanewala~\cite{rahman2018predicting} and Nair et al.~\cite{nair2019leveraging}.


Zeng et al.~\cite{jin2016multi} proposed a neural network based approach to MR prediction. Similar to previous studies~\cite{kanewala2013using, kanewala2014techniques}, the proposed approach also used CFG to extract features, which, in turn, can lead to a multi-label dataset. However, instead of the machine learning based supervised or semi-supervised classifiers~\cite{kanewala2016predicting, hardin2018using}, this approach applied the radial basis function (RBF) neural network to obtain the predictive model. 
Other studies on MR prediction include those based on program documentation~\cite{rahman2020mrpredt} and using the semi-supervised SVM-bagging KNN algorithm~\cite{yin2021syntax}.

Different from MR prediction techniques that require pre-defined relations, a recent study~\cite{MRgpt23} leveraged a large language model (LLM) to automatically generate MRs for autonomous driving. More specifically, ChatGPT (GPT-3.5) was asked to generate a certain number of MRs for the parking module in an autonomous driving system. It was found that the relations generated by ChatGPT had varying quality, dependent on various factors, including how complex the system under test was, how specifically the test scenario was defined, and even how many MRs ChatGPT was asked to generate. Human interventions, either pure manual checking or more interactions with ChatGPT, were required to validate whether a generated relation was a correct MR and/or to improve the quality of the generated MRs. Such LLM-based techniques are particularly useful for the so-called end-user software development~\cite{EUSE11}, where programs are not written by the professional software engineers but by the end users with expertise in other areas. MT has been recognized as an effective technique to support the end-user testing~\cite{eutest05}. The use of LLM makes it easier for end users to generate MRs even without in-depth knowledge and extensive experience of MT. Moreover, the domain expertise of end users will be of much help in the validation and selection of appropriate MRs for implementing MT.

\subsection{Metamorphic Relation Patterns}

In early MT research, MRs were mostly identified from scratch. To make this process more systematic, Zhou et al. were the first to present an idea of using an \textit{abstract} form of MRs, which they called a ``general metamorphic relation''~\cite{Zhou07HKU}, to derive multiple \textit{concrete} MRs. For example, in the context of testing search engines, they defined a general MR $\textit{MR}_\textit{SEARCH}$ as follows:
\begin{quote}
	$\textit{MR}_\textit{SEARCH}$: Let $X$ be a search criterion and \#($X$) be the number of pages returned for search criterion $X$. A \textit{general metamorphic relation} is: if $X_2$ implies $X_1$, then \#($X_2$) $\leq$ \#($X_1$).
\end{quote}
Then, from $\textit{MR}_\textit{SEARCH}$, they derived multiple concrete MRs for the ``AND,'' ``OR,'' and ``EXCLUDE'' search operators.
In a follow-up study, 
Zhou et al. 
further identified another type of abstract relation, namely, a \textit{subset} relation among the source and follow-up \textit{outputs}~\cite{Zhou12automated}. 

When studying MT for RESTful Web APIs, Segura et al.~\cite{segura2017metamorphic} defined a term \textit{metamorphic relation output pattern} (MROP) as ``an abstract output relation typically
identified in Web APIs, regardless of their application domain.'' An MROP relates the source and follow-up \textit{outputs} without putting any restriction on the relations among multiple inputs.
Segura et al. identified six MROPs in terms of set operations for RESTful Web APIs, including
equivalence, equality, subset, disjoint, complete, and difference.
Their work opened a new MT research direction for ``metamorphic relation patterns'' in a broad sense.
The same research team studied the inference of likely MRs for the verification of model transformations~\cite{Troya18}. They first identified a set of situations (\textit{domain-independent trace patterns}) that may occur during a model transformation. Then, for each of these situations, they defined a set of likely (domain-independent) MRs that may be instantiated to become more concrete MRs when the context of a specific model transformation is given.


Zhou et al.~\cite{ZhouTSE19} further studied the notion of ``patterns'' and formally defined the general concept of \textit{metamorphic relation pattern} (MRP) as ``an abstraction that characterizes a set of (possibly infinitely many) metamorphic relations.'' 
It was pointed out that a collection of patterns can form a hierarchy, with those at higher levels being more abstract. At the top level of the MRP hierarchy, they identified a \textit{symmetry} MRP, which refers to the ``existence of different viewpoints from which the system appears the
same.'' It should be noted that ``the system appears the same'' does not
mean that the source and follow-up outputs must have an equality relation.
Zhou et al.~\cite{ZhouTSE19} provided examples and case studies 
to show that concrete instances derived from the symmetry MRP are almost everywhere. They hypothesized that \textit{symmetry} and \textit{asymmetry} are two fundamental MRPs that come in pairs for software systems.

Due to the wide diversity of application domains, it is extremely difficult, if not impossible, to find MR patterns that are generally applicable to different types of software. Therefore, the most recent studies in this context were focused on designing specific MR patterns for certain domains. For example, Sun et al.~\cite{ConMT23} made use of 14 interleaving patterns to design MRs specifically for the testing of concurrent programs. Liu et al.~\cite{DiffFuzz23} defined the so-called ``symmetry MR pattern'', based on which a series of MRs were constructed for the fuzz testing of deep learning libraries. 

\subsection{Category-Choice Framework Based Techniques}
\label{sec:metric}


The category-choice framework is based on the {\sc C}ategory-{\sc P}artition {\sc M}ethod (CPM)~\cite{Ostrand88} and the {\sc CHO}i{\sc C}e-re{\sc LAT}ion fram{\sc E}work ({\sc CHOC'LATE})~\cite{chen2003choice,poon2010choc}. In this framework, categories and their corresponding choices are defined from a software specification. Here, a \textsl{category} $X$ is a property or characteristic of an input parameter or environment condition, where \textsl{choices} of $X$ are the non-overlapping subsets of the values of $X$. \textsl{Complete test frames} (CTFs) are then generated based on the valid combinations of choices from different categories. If a value is selected from every choice of a CTF, a test case will be formed. Thus, the category-choice framework was originally developed as a test case generation method.

Chen et al.~\cite{chen2016metric} observed that MRs can be constructed by comparing two different test scenarios (each scenario refers to a set of test inputs) and predicting the relations among the corresponding outputs (if any). This can be achieved by using the category-choice framework because each CTF is effectively a test scenario. Based on this notion, they proposed the {\sc METRIC} technique~\cite{chen2016metric}. In {\sc METRIC}, a \textsl{candidate pair} is first formed by selecting two relevant and distinct CTFs (represented by CTF$_1$ and CTF$_2$). The tester then manually evaluates whether the candidate pair is useful for MR construction. In the evaluation process, the tester asks the question: Given any two inputs $x$ and $y$ (where $x$ is any input constructed from CTF$_1$ and $y$ is any input from CTF$_2$), is there a definite relationship between the outputs corresponding to $x$ and $y$? If yes, CTF$_1$ and CTF$_2$ form a \textsl{usable pair} because an MR can be defined from them. In this case, the tester continues to define the MR description corresponding to CTF$_1$ and CTF$_2$. This selection-evaluation-construction process will continue until a pre-set termination condition is met, e.g., all the candidate pairs relevant to the selection criteria are exhausted, or the preferred number of MRs to be generated (which can be initially defined by the tester before the process starts) is reached. 

Sun et al.~\cite{sun2019metric} extended the original {\sc METRIC} technique by considering both the input and output domains of the software under test, resulting in a new technique called {\sc METRIC}$^{+}$. In {\sc METRIC}$^{+}$, the ``original'' categories and choices solely defined based on the input domain were called \textsl{I-categories} and \textsl{I-choices} (``I'' stands for ``Input''), respectively. Additionally, two new sets of concepts were used in {\sc METRIC}$^{+}$: ($a$) \textsl{O-categories} and \textsl{O-choices} (``O'' stands for ``output''), which are defined based on the output domain only, and ($b$) \textsl{IO-based complete test frames} (IO-CTFs) (``IO'' stands for ``Input and Output''), each of which is a valid combination of I-choices and O-choices. With this extension, more guidelines can be given for finding the relationship between outputs (leveraging the O-categories and O-choices), instead of pure manual evaluation for output sub-relation in {\sc METRIC}. 

Intuitively speaking, the category-choice framework can be applied into any type of software provided that categories and choices can be defined. Therefore, {\sc METRIC} and {\sc METRIC}$^{+}$ have very wide applicability, i.e., they are not specific to certain types of software.

\subsection{Genetic Approach}
Ayerdi et al.~\cite{ayerdi2021generating} proposed a technique for constructing MRs automatically by means of genetic programming~\cite{terragni2021improving,terragni2020evolutionary,terragni2021gassert}. The method first samples some correct and incorrect executions of the system under test, and then uses them as a basis to formulate MR generation and improvement as a multi-objective optimization problem with three objectives: ($a$) minimizing the number of false positives, ($b$) minimizing the number of false negatives, and ($c$) minimizing the number of generated MRs.
The method can only handle numeric inputs and outputs, and the generated MRs cannot involve multiple outputs. In view of this limitation, if MRs were required for multiple outputs, the method must be executed multiple times, with each execution generating MRs for one particular output.

Xiang et al.~\cite{xiang2019genetic} have also worked on MR generation using the genetic approach. Their method, however, differed from the one by Ayerdi et al.~\cite{ayerdi2021generating} in that the former~\cite{xiang2019genetic} focused on MR composition and, hence, required the existence of some ``initial'' MRs. More specifically, Xiang et al.~\cite{xiang2019genetic} used a genetic approach for generating multi-layer composite MRs. Their approach considers composite MR construction as a problem for searching optimal composite sequences, which is solved by an adaptive heuristic optimization algorithm based on natural selection and genetic evolution~\cite{holland1973genetic}. In general, the genetic approach proposed by Xiang et al.~\cite{xiang2019genetic} involves the following two major steps: ($a$)~Initialize individuals of composite MR sequence (each composite MR sequence is a possible solution to the problem) within the search space; ($b$)~Iteratively determine the fitness value for each individual of composite MR sequence in the population that satisfies $R_i(I_1, I_2) \Rightarrow R_o(O_1, O_2)$. A new population is then constructed by evolutionary operators (i.e., selection, crossover, and mutation) based on the fitness value. Step~($b$) will be repeated for several times, and the individual composite MR sequence with the best overall fitness value will be returned.

Hong et al.~\cite{MRgep22} proposed a technique to ``recognize'' MRs based on input patterns and gene expression programming (GEP -- a swarm intelligence evolutionary algorithm). The method is only applicable to scientific programs with numeric inputs. It first pre-sets some input relations manually, based on certain patterns. These input relations are then used to generate follow-up inputs based on randomly generated source inputs. Possible output relations will be mined by GEP between source and follow-up outputs after executing the program under test with all inputs. The mined output relations will be reduced and validated. After validation, the input relation and output relation can compose an MR.

\subsection{Search-Based Techniques}
\label{sec:search}

The first search-based MR generation method was proposed by Zhang et al.~\cite{zhang2014search}. The method is only applicable to programs with numeric inputs and outputs, and can only generate MRs with separate input and output sub-relations. It can automatically construct the so-called ``polynomial MRs'', where the input sub-relations are restricted to linear expression of source and follow-up inputs, while the output sub-relations are restricted to be either linear or quadratic expressions of source and follow-up outputs. More specifically, a popular search-based technique, particle swarm optimization (PSO), is leveraged to predict coefficients for both the input and output expressions based on input-output pairs. The method is basically a black-box approach, as it only refers to the program execution without any in-depth consideration of the source code. However, it should be noted that the relations generated by the method may not be the valid MRs, because a wrong conclusion may be drawn with limited pairs of inputs and computed outputs, or because the program being used is incorrectly implemented. In order to enhance the quality of the generated MRs, Zhang et al. applied a statistics-based filtering process to remove the ``low-quality'' MRs and only accept those MRs that are satisfied by a huge amount of randomly generated inputs. Their experimental studies showed that PSO did help to generate high-quality MRs.

Zhang et al.~\cite{zhang2019automatic} extended the above PSO-based approach~\cite{zhang2014search} and proposed a new method called AutoMR. AutoMR improved the diversity of the generated MRs: It can generate MRs that involve more than two inputs, and extend from equality relations only to both equality and inequality relations. In addition, AutoMR makes use of constraints solving and singular value decomposition to remove some redundant MRs. As compared with the original work~\cite{zhang2014search}, AutoMR~\cite{zhang2019automatic} also uses PSO to search for the coefficients of the polynomial expressions, but it utilizes a more fine-grained fitness function, resulting in MRs of higher quality.

\subsection{Miscellaneous}
\label{sec:others}
In addition to the representative methods discussed in Sections~\ref{sec:composition} to~\ref{sec:search}, studies have been conducted to develop many other techniques for generating MRs systematically. The following gives some examples of these techniques:

\textbf{Data mutation}. Sun et al.~\cite{sun2016mumt, muData24} proposed a data mutation directed MR generation approach called $\mu$MT. Data mutation (DM)~\cite{dataMut06} is a test case generation technique inspired by mutation testing. The $\mu$MT approach uses DM operators to identify relations among inputs, and some mapping rules of the program under test are utilized to identify the output relations of the related test cases. 

\textbf{Iterative development}. Ding and Zhang~\cite{ding2016approach} introduced the idea of ``iterative refinement'' for MR generation. Specifically, five initial MRs were manually defined for a Monte Carlo simulation program. Some initial test cases were generated to further understand the MRs. The approach visualized the input-output relations via some curves connecting the input-output points. By observing these points and curves, testers would manually figure out some problems with the program under test or the used MRs. By adding more points into the curves, more precise information would be obtained to understand the properties of the target program. Thus, the MRs can be refined, e.g., 
``$x > y$'' $\to$ ``$x-y = 10$''.
Such a process can be repeated until the testing is ``adequate'', which was judged by some criteria, such as coverage.

\textbf{Natural language processing (NLP)}. Blasi et al.~\cite{blasi2021memo} presented a technique, MeMo, for automatically retrieving equality MRs from natural language documents. They argued that a development document with complete and correct code comments may contain many important keywords, such as ``similar'', ``like'', and ``same as'', which describe the equality relations. Accordingly, the MeMo approach consists of the following steps: (1) The Comments Processor removes useless information and splits the cleaned comments of each class into sentences for further MR findings;  (2) The MR Finder determines whether or not each sentence describes an equality MR by checking if it contains one of the equivalent keywords; (3) The Translator converts all sentences marked as describing equality relations into MRs as Java assertions; (4) The Executor verifies whether or not the Java assertions are valid against the source code.

\textbf{Template-Based}. Segura et al.~\cite{segura2017template} proposed a template-based approach for describing MRs, where a template refers to ``a combination of placeholders and linguistic formulas used to describe something in a particular domain''. Segura et al. attempted to design the template to be ``intentionally simple and flexible'', based on MRs ``observed in the literature''. They further provided a number of examples by using the template to describe MRs identified in other studies~\cite{chen2016metamorphic,lindvall2016agile,Zhou16,xie2011testing, segura2015automated}. 
They also validated the approach through applying it to describe 17 randomly selected MRs from ten MT papers (involving 35 authors and eight different applications domains)~\cite{segura2017metamorphic}, and no expressiveness problems were found for the MRs described with the template.
Some research directions were pointed out, including a more rigorous validation as well as the mapping of template-based MRs to executable code or test cases~\cite{segura2017template}.

\textbf{Guideline based}. Raunak and Olsen~\cite{raunak2021metamorphic} investigated how to generate MRs for verifying and validating simulation models. They differentiated MRs for testing simulation models into two categories, those for verification and those for validation. They provided a guideline on how to define these two types of MRs. In the guideline, after the non-testable ``aspects'' are identified from the simulation model, for each aspect, the tester should consider how a change of a parameter or a change to the model (i.e., input relation) will affect the system. If one can predict the change in result (i.e., output relation) from expert domain knowledge, an MR for validation is then defined. If one can predict the change in result from the code or implementation details, an MR for verification is then defined. 

\textbf{Tabular expression}. Li et al.~\cite{li2020tabular} proposed an approach to MR construction based on specifications written in tabular expressions. The features of tabular expressions were leveraged to defined 11 ``general MRs'', based on which more ``concrete'' MRs can be generated. Among these MRs, six MRs involved one source test case and one follow-up test case, while the other five MRs involved multiple source and multiple follow-up test cases. These general MRs were constructed based on the so-called ``normal form'' cell connection graph (CCG). 

\textbf{Behavior driven}. Deng et al.~\cite{deng2021bmt} proposed a declarative behavior-driven development (BDD)-based MT technique (BMT) for the testing of autonomous driving systems. BMT enables descriptions and specifications of customized traffic behavior, then translates the behavior into an MR, and ``synthesizes'' meaningful test inputs using imaging and graphics processing techniques. They extended this work to a more comprehensive declarative rule-based MT framework (RMT)~\cite{RMTTSE23}, which includes an NLP-based semantic parser for extracting critical information from traffic rules and then automatically creating corresponding MRs.

\textbf{Online game}. Yang and Xu~\cite{yang2020mr} presented how to hypothesize and validate MRs in a puzzle-solving context using an interactive mode. They developed a platform called M.R. Hunter that is particularly suitable for crowd sourcing paradigm as well as for training and learning. The main idea is to hypothesize an MR from given input-output pairs. Technically speaking, testers may suggest some inputs for the program to execute to get their corresponding outputs. With these input-output pairs, testers then hypothesize a potential MR which could be further validated by means of executing some additional inputs. As an example of illustration, from a set of input-output pairs, suppose a tester has hypothesized an MR that if the input is negated, its output will also be negated. In order to validate this potential MR, the tester may negate some of the already executed test cases to see whether or not the expected outputs would be returned. If yes, the hypothesized MR is confirmed; otherwise, the hypothesized MR is deemed invalid. 

\textbf{User forum}. Lin et al.~\cite{lin2021finding} presented an approach for generating MRs from user forums of a scientific simulation system. In their approach, machine learning techniques are leveraged to locate and classify variables from user forums. The association of inputs and outputs are then mined to obtain the candidate MRs based on classified variables. Finally, they validated the candidate MRs through regression tests, as well as ranking MRs according to the confidence that determines the relative number of an outcome among all alternatives for a given antecedent condition. 

\textbf{Hierarchical}. Lin et al.~\cite{lin2018hierarchical} proposed the so-called ``hierarchical MRs'' for scientific software with multiple parameters. In their approach for MR generation, they considered the calibrations of different parameters and their influences on the system output. On the first level, MR$_{sp}$ (Singleton versus Pair), the implementation of a pair of parameters is compared with the separate implementations of single parameters. In other words, there are two source test cases, each referring to one parameter, while the one follow-up test case refers to the pair of the corresponding two parameters. If the calibrations of the pair and singletons have a certain relation (e.g., joint calibration is better than the calibration on either single parameter), an MR is then defined. On the second level, MR$_{pt}$ (Pair to Triplet), the source test case involves a pair of parameters, while the follow-up test case has a third parameter. If the third parameter makes the calibration better, an MR is then defined. The third level, MR$_{FL}$ (Fault Localization), as the name states, is supposed to help locate the fault. For example, if the third parameter in the MR$_{pt}$ is altered but the output remains unchanged, a fault is said to be located in the model. This can then be considered as a new MR. 

\section{New Visions}
\label{sec:ResearchTrends}
\begin{figure}[t]
	\centering
	\includegraphics[width=\textwidth]{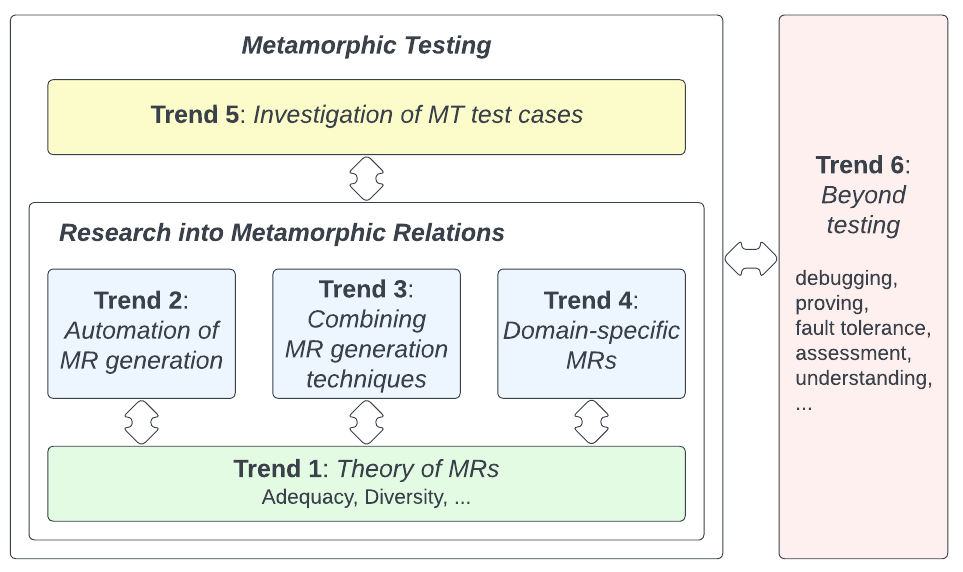}
	\caption{Future research trends}
	\label{fig:Research_trend}
\end{figure}

In this section, we present six research trends that we consider the most promising for the future work. As summarized in Figure~\ref{fig:Research_trend}, the first four trends are directly related to MR generation, while the last two go beyond MRs or even testing.

\vspace{3pt} \noindent \textbf{Trend 1: Theory of MRs}

Although various types of experimental studies have been conducted on MRs from different perspectives, there exists much less work to establish a theoretical foundation of MRs and hence MT. We hereby recommend two research directions about the theory of MRs, namely the adequacy and diversity of MRs.

Generally speaking, an application may have a huge number of necessary properties, which can be converted to MRs. Analogous to the testing principle ``exhaustive testing is impossible'', it is also impossible to exhaustively generate all possible MRs for a target system. It is thus critical to establish the knowledge on judging whether and to what extent the generated MRs are ``sufficient'' for MT. By nature, there are different perspectives to interpret a program, such as the black-box requirements specification and the white-box code structure. A certain perspective can, in turn, be denoted by a set of characteristics, such as the specification conditions and structural elements.  Intuitively speaking, a set of MRs can be considered as effective if they can thoroughly cover certain characteristics of the software under test, which implies the comprehensive coverage of the software functionalities and execution behaviors, and thus a high potential of detecting most faults. In other words, the adequacy of MRs can be judged through considering various aspects of software characteristics.

As more and more MR generation techniques are being developed, we are experiencing a rapidly increasing number of MRs for various application domains. This inspires another question: Can we just use a subset of MRs without jeopardizing the fault-detection effectiveness? The answer we have learned from previous empirical studies~\cite{Liu14TSE} is the diversity of MRs, that is, they should be able to trigger as different functionalities and execution behaviors as possible, which can be reflected by as different characteristics covered by MRs as possible. Investigations can be done to define a set of formulas and metrics based on certain software characteristics for evaluating the diversity of MRs from different perspectives. 

Integrating the concepts of adequacy and diversity with the systematic MR generation techniques will result in a series of advanced techniques that can achieve the high effectiveness and high efficiency, or simply high cost-effectiveness, of testing. 

\vspace{3pt} \noindent \textbf{Trend 2: Automation of MR generation}

Automation of everything is one dream of many software engineers. The proposal of systematic MR generation techniques can facilitate the achievement of such a dream, at least partially. However, it should be pointed out that human intelligence is still a critical part --- some manual work is required to derive some entities (e.g., the categories and choices for {\sc METRIC}/{\sc METRIC}$^{+}$~\cite{chen2016metric, sun2019metric} and pre-defined relations for MR prediction~\cite{kanewala2016predicting}) as the basis for generating MRs. In other words, it may not be feasible to achieve a full automation of MR generation, but we should automate the process as much as possible.

Take the {\sc METRIC} and {\sc METRIC}$^{+}$ techniques as an example. When a tester applies {\sc METRIC}, he/she must first derive categories and choices, mostly in a manual manner. Some tools can automatically suggest candidate pairs of test frames, on which the tester should, again, manually determine whether each pair can have a certain relation. After the determination based on the tester's human intelligence, a set of MRs can be constructed and expressed in a pre-defined format. As an enhanced version of {\sc METRIC}, {\sc METRIC}$^{+}$ can improve the degree of automation. It leverages the output-related categories and choices to give more clues for determining the potential relation of a candidate test frame pair. In other words, a {\sc METRIC}$^{+}$-based tool can further automatically suggest ``potential'' relations for the tester to verify or further polish. 

It is intuitively expected that most proposed MR generation techniques can be automated to some extent, as long as they contain some systematic mechanism for a machine to follow in a step-by-step way. Nevertheless, the current practices suggest that the human intelligence is still essential for MR generation.

\vspace{3pt} \noindent \textbf{Trend 3: Combining different MR generation techniques}

Different MR generation techniques have different advantages and disadvantages. Since many of them are not mutually exclusive to each other, it is natural to investigate their combinations to develop more advanced MR generation techniques.

For example, {\sc METRIC}/{\sc METRIC}$^{+}$~\cite{chen2016metric, sun2019metric} can construct a large number of MRs. The MT composition technique can combine existing ``component'' MRs to generate new ``composite'' MRs: The latter can ``inherit'' features of the former, and thus a smaller set of composite MRs can achieve similar fault-detection effectiveness to the full set of component MRs~\cite{qiu2020theoretical}. One straightforward way is to utilize the MRs constructed by {\sc METRIC}/{\sc METRIC}$^{+}$ as the basis for MR composition. In this way, we can obtain a relatively small number of MRs without jeopardizing the fault-detection effectiveness, thus improving the cost-effectiveness of MT. Another possible combination is the use of MR prediction techniques~\cite{kanewala2016predicting} to validate whether or not the relations generated by other methods (e.g., those given by an LLM~\cite{MRgpt23}) are correct MRs.

\vspace{3pt} \noindent \textbf{Trend 4: Domain-specific MRs}

Many MR generation techniques are applicable to certain domains, e.g., those for object-oriented programs~\cite{zhang2016method}, cybersecurity~\cite{ChaleshtariTSE23}, and concurrent programs~\cite{ConMT23}. It has been consistently shown that domain-specific MRs are particularly effective in detecting faults in the corresponding areas. This is understandable, as the specific features (here represented by MRs) of a certain domain should have strong correlation to the faults of specific types in the software of that domain. Therefore, it is worthwhile to explore domain-specific techniques for constructing MRs in different areas, for which we suggest the following three research directions: ($i$) Design of specific MR patterns for certain domains, like the work for search engines~\cite{Zhou12automated} and deep learning~\cite{DiffFuzz23}. ($ii$) Development of domain-specific languages to describe MRs. For example, Chaleshtari et al.~\cite{ChaleshtariTSE23} proposed the Security Metamorphic Relation Language (SMRL) to present MRs they have derived for the security testing of Web systems. On this research direction, we can investigate how to specify MRs formally for a certain application domain such that they can be automatically converted to executable code, thus making the automation of MT easier~\cite{ChaleshtariTSE23, RMT23}. ($iii$) Proposal of new concepts of MRs. Sun et al.~\cite{ConMT23}, for instance, proposed that MRs for the testing of concurrent programs should be augmented with ``interleaving scenarios'' to improve the effectiveness of detecting interleaving-related concurrency bugs, resulting in a new notion of MR. In the future, the similar work can be done for different application domains: We should fully consider the specific features of a certain domain when generating MRs.

\vspace{3pt} \noindent \textbf{Trend 5: Investigation of MT test cases}

As demonstrated by numerous studies, MRs have helped generate new test cases that were able to reveal long-standing bugs. This implied that beyond alleviating the oracle problem, MRs could also be considered as a practical means to ``top up'' test case generation (that is, generating test cases supplementary to those generated by other traditional testing techniques). Some initial studies have been conducted to compare the fault-detection effectiveness of follow-up test cases and their corresponding source test cases~\cite{followBetter21}. It is worthwhile to conduct comprehensive investigations on the effectiveness of MT test cases, especially in comparison with the counterparts created by the traditional testing techniques. The effectiveness of MT mainly depends on MRs and source test cases~\cite{chen2018metamorphic}. Some advanced testing techniques~\cite{Alatawi16, MTartMET16, FDMTTOSEM23} have been used to generate ``good'' source test cases for MT. More research should be conducted on this direction such that MT's performance can be maximized by using both good MRs and source test cases.

Another important research direction is the identification of the actual failure-revealing test cases in MT. When an MR is violated, we can say that a failure is revealed by multiple test cases (including both source and follow-up test cases). However, it is still unknown which test case(s) on earth are actually responsible for the revealed failure (although some test cases are involved in an MR violation, they may not be failure-revealing). In many other software quality assurance activities, such as debugging and fault tolerance, such precise information of failure-revealing test cases is very important for them to make the accurate decisions. The research on finding the actual failure-revealing test cases will help improve the applicability of MT in other areas.

\vspace{3pt} \noindent \textbf{Trend 6: Beyond testing}

The basic idea of MT has been applied into various areas beyond testing, including debugging~\cite{Xie13}, proving~\cite{Chen10Semi}, fault tolerance~\cite{MTFTICSE14}, system assessment~\cite{Zhou16} and understanding~\cite{ZhouTSE19}. These applications actually are based on the usage of MRs in the relevant fields; in other words, MRs have evolved beyond software testing, or even verification and validation. We envisage that new roles and uses of MRs will continue to emerge, thus posing new requirements on MR generation. In the context of validation, MRs should be generated by fully considering the expectations from the users' perspective. In the context of systems assessment or selection, we should focus on generating MRs that can serve as evaluation, adequacy, and appropriateness criteria. In the context of software understanding, we should generate MRs that can facilitate the user's comprehension in the absence of a thorough specification. Going beyond testing, some research can be done to generate MRs that are suitable for these new roles and uses.

\section{Conclusion}
\label{sec:conclusion}

Since its invention in the 1990s, metamorphic testing (MT) has been applied into the testing of various types of software systems, resulting in great successes of detecting critical real-life bugs in a wide variety of application domains. Metamorphic relations (MRs), the core component of MT, have been recognized as the key factor for these successes, and thus have attracted a lot of research interest. Particularly, in recent years, the community has experienced a rapidly increasing number of investigations into the techniques of generating MRs systematically, which is the focus of this article. We have identified and reviewed the mainstream MR generation techniques, including those based on composition, artificial intelligence, category-choice framework, patterns, genetic and search algorithms. We also discussed some ``miscellaneous'' techniques from different perspectives, such as natural language processing and data mutation. Finally, we highlighted the potential directions that require special attentions for the research in MRs and more broadly, MT, ranging from establishing foundational theory of MRs to developing more advanced MR generation techniques, and from investigating MT test cases to applying MRs and MT into other software quality assurance activities. 

\section{Acknowledgments}
This work was supported by the Australian Research Council Project under Grant DP210102447.

\end{document}